\documentclass[amsmath,amssymb,aps,pra,twocolumn]{revtex4-2}
\usepackage{graphicx}
\usepackage{bm}
\usepackage[dvipdfmx]{hyperref}
\usepackage{float}
\begin{document}

\title{Thermodynamics of a continuous quantum heat engine: 
Interplay between population and coherence}
\author{Pablo Bayona-Pena}
\affiliation{Department of Physics, Tokyo Institute of Technology, Tokyo 152--8551, Japan}
\author{Kazutaka Takahashi}
\affiliation{Institute of Innovative Research, Tokyo Institute of Technology, Kanagawa 226--8503, Japan}

\date{\today}

\begin{abstract}
We present a detailed thermodynamic analysis of a three-level
quantum heat engine coupled continuously to hot and cold reservoirs.
The system is driven by an oscillating external field
and is described by the Markovian quantum master equation.
We use the general form of the dissipator 
which is consistent with thermodynamics. 
We calculate the heat, power, and efficiency of the system 
for the heat-engine operating regime 
and also examine the thermodynamic uncertainty relation.
The efficiency of the system is strongly dependent on the structure of
the dissipator, and the correlations between different levels can be 
an obstacle for ideal operation.
In quantum systems, the heat flux is decomposed into 
the population and coherent parts. 
The coherent part is specific to quantum systems, and 
in contrast to the population part, 
it cannot be expressed by a simple series expansion 
in the linear-response regime.
We discuss how the interplay between the population and coherent parts 
affects the performance of the heat engine.
\end{abstract}
   
\maketitle

\section{Introduction}

One of the main objectives in the field of quantum thermodynamics is 
to find the quantum signatures of heat engines. 
The thermodynamical description is applied even when the system
has a few degrees of freedom, which allows us to study
the quantum effects on heat and work from a microscopic point of view.

Since the seminal work by Scovil and Schulz-DuBois~\cite{Scovil59},
the microscopic heat engines with a few discrete energy levels 
have been studied in many 
works~\cite{Kosloff84,Geva94,Geva96,Boukobza06,Boukobza07,Uzdin15,Singh20}.
An upper bound on the efficiency of the heat engine is given
by the Carnot efficiency, as expected from the general
arguments~\cite{Spohn78,Alicki79}.
Experimental examples of quantum heat engines have been realized
in a wide range of quantum systems such as trapped ions and
ensembles of nitrogen-vacancy centers
in diamond~\cite{Rossnagel16,Maslennikov19,Klatzow19}.

It is an important problem to answer the question of whether 
the quantum coherence enhances the power output of the heat engine.
Several works have shown that the answer is
positive~\cite{Scully03,Scully11,Rahav12,Uzdin15},
which represents a promising feature for the design of microscopic devices.

Generally speaking, quantum effects arise 
when the density operator has off-diagonal components in
the basis of the Hamiltonian operator.
The quantum coherence is characterized by the off-diagonal parts, and
the nonequilibrium entropy production can be decomposed into
the classical population part and the quantum coherent
part~\cite{Baumgratz14,Santos19}.
The Markovian dynamics is described by the 
Gorini--Kossakowski--Lindblad--Sudarshan (GKLS)
equation~\cite{Gorini76,Lindblad76,Breuer02}.
By decomposing the dissipator into two parts,
we can introduce the corresponding heat for each part~\cite{Funo19}.

In the present work, we revisit the three-level quantum heat engine
developed by Geva and Kosloff~\cite{Geva94}
in order to examine the roles of the population and 
coherent parts of the heat flux.
We also study how the result depends on the details of 
the dissipator part of the GKLS equation.
The choice of the dissipator is important 
for finding thermodynamically consistent results, 
and we treat a possible general form of the dissipator.

This paper is organized as follows.
In Sec.~\ref{sec:system}, we describe the settings of our model and
define the heat, work, and efficiency according to the standard scenario.
Section \ref{sec:pandeff} summarizes 
the stationary solution of the model.
We obtain the explicit forms of the heat, power, and efficiency.
In Sec.~\ref{sec:decomp}, we introduce the concept of population heat 
currents and coherent heat currents to discuss how
the quantum nature affects the performance.
We also discuss the thermodynamic uncertainty relation (TUR) in 
Sec.~\ref{sec:tur}.
The conclusion is summarized in Sec.~\ref{sec:conc}.

\section{System setting and energetic relations}
\label{sec:system}

\subsection{GKLS equation}

\begin{figure}[t]
\includegraphics[width=0.80\columnwidth]{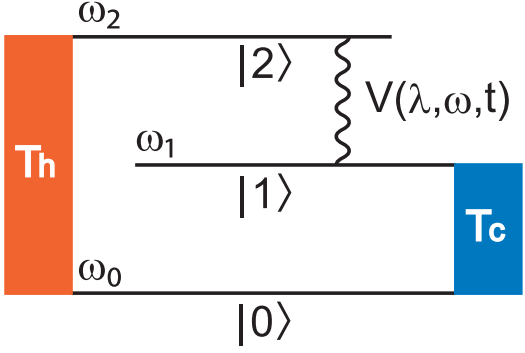}
\caption{A schematic view of the three-level quantum heat engine.
The system is driven by the time-dependent field
$V(\lambda,\omega,t)=\lambda e^{i\omega t}|1\rangle\langle 2|
+\lambda e^{-i\omega t}|2\rangle\langle 1|$.
}
\label{fig01}
\end{figure}

We define a continuous quantum heat engine 
by using a three-level maser as shown in Fig.~\ref{fig01}.
The system consists of three states $|0\rangle$, $|1\rangle$, 
and $|2\rangle$, 
and each level has energy $\omega_0$, $\omega_1$, and $\omega_2$,
respectively. 
We set $\omega_0<\omega_1<\omega_2$.
Two of the states, $|0\rangle$ and $|1\rangle$, are coupled to 
the cold reservoir with temperature $T_{\rm c}=1/\beta_{\rm c}$,  
and $|0\rangle$ and $|2\rangle$ are coupled to the hot reservoir 
with $T_{\rm h}=1/\beta_{\rm h}$, where $T_{\rm h}> T_{\rm c}$.
We study the performance of the heat engine when 
the system is continuously coupled to the two heat reservoirs.
The system is operated by applying 
a time-oscillating field with frequency $\omega$.
The field induces transitions between $|1\rangle$ and $|2\rangle$,
generating a heat flow, as we discuss below.
The system Hamiltonian is given in the state basis as 
\begin{equation}
 \hat{H}(t) = \begin{pmatrix}
 \omega_0 &0&0 \\
 0&\omega_1 & \lambda e^{i\omega t}\\
 0 & \lambda e^{-i\omega t}&\omega_2 
 \end{pmatrix}, \label{hamiltonian}
\end{equation}
where $\lambda$ represents the field intensity.
This Hamiltonian is diagonalized as
$\hat{H}(t)=\sum_{n=0}^2\epsilon_n|\epsilon_n(t)\rangle\langle\epsilon_n(t)|$, 
and the energy eigenvalues are independent of $t$, 
as shown in Appendix~\ref{app-eigen}.
We assume $\lambda^2< \omega_{10}\omega_{20}$, 
where $\omega_{10}=\omega_1-\omega_0$ and $\omega_{20}=\omega_2-\omega_0$, 
so that the order of the energy levels is unchanged, 
$\epsilon_0<\epsilon_1<\epsilon_2$, in the presence of the field.

The present setting is basically the same as discussed in related 
works~\cite{Geva94,Geva96,Boukobza06,Boukobza07,Singh20}.
A difference arises when we carefully treat the dissipation effect.
The state of the system is described by the density operator 
$\hat{\rho}(t)$.
We assume the Markovian dynamics, and 
the time evolution is described by the GKLS equation,
\begin{equation}
 \partial_t\hat{\rho}(t)= -i[\hat{H}(t),\hat{\rho}(t)]
 +\sum_{\alpha={\rm c},{\rm h}}\hat{D}_\alpha[\hat{\rho}(t)].
 \label{gkls}
\end{equation}
The dissipation affects the quantum dynamics in three ways.
Two of them are described by the dissipator $\hat{D}_\alpha$.
It changes the eigenbasis of the density operator and 
the population of each basis element~\cite{Funo19}.
The remaining one shifts only the energy levels of 
the Hamiltonian $\hat{H}(t)$ and is called the Lamb shift.
Here we simply neglect the Lamb shift term or 
take $\hat{H}(t)$ to be a renormalized Hamiltonian.

The dissipator $\hat{D}_\alpha$ is chosen so that the time evolution 
is a completely positive and trace-preserving map.
The explicit form of the dissipator 
is dependent on coarse-graining procedures.
In order to obtain a thermodynamically consistent description 
for the present problem, we use the form 
\begin{eqnarray}
 &&\hat{D}_\alpha[\hat{\rho}] = \sum_{\epsilon}\gamma_{\alpha}(\epsilon)
 \biggl[
 \hat{L}_{\alpha}^{\epsilon}(t)\hat{\rho}(\hat{L}_{\alpha}^{\epsilon}(t))^\dag
 \nonumber\\
 && -\frac{1}{2}\left( 
 (\hat{L}_{\alpha}^{\epsilon}(t))^\dag \hat{L}_{\alpha}^{\epsilon}(t)\hat{\rho}
 +\hat{\rho}(\hat{L}_{\alpha}^{\epsilon}(t))^\dag \hat{L}_{\alpha}^{\epsilon}(t)
 \right)
 \biggr], \label{dissipator}
\end{eqnarray}
where $\hat{L}_{\alpha}^{\epsilon}(t)$ represents projected jump operators 
\begin{eqnarray}
 && \hat{L}_{\alpha}^{\epsilon}(t) = (\hat{L}_{\alpha}^{-\epsilon}(t))^\dag \nonumber\\
 &&=\sum_{m,n=0}^2\delta_{\epsilon,\epsilon_m-\epsilon_n}
 |\epsilon_n(t)\rangle\langle \epsilon_n(t)|
 \hat{L}_{\alpha}|\epsilon_m(t)\rangle\langle \epsilon_m(t)|,
\end{eqnarray}
with
\begin{eqnarray}
 && \hat{L}_{\rm c}=|0\rangle\langle 1|, \label{lc}\\
 && \hat{L}_{\rm h}=|0\rangle\langle 2|. \label{lh}
\end{eqnarray}
The dissipator coupling 
$\gamma_{\alpha}(\epsilon)$ is generally nonnegative and
is a system-dependent function of $\epsilon$ and $\beta_\alpha$.
To construct a thermodynamically consistent theory,
we assume the detailed balance condition 
\begin{equation}
 \gamma_{\alpha}(-\epsilon)=  e^{-\beta_{\alpha}\epsilon}\gamma_{\alpha}(\epsilon). 
 \label{dbc}
\end{equation} 
Then, the Gibbs distribution becomes 
an instantaneous stationary solution~\cite{Breuer02,Funo19}.

We can derive the form of the dissipator in Eq.~(\ref{dissipator})
from a microscopic model by using several approximations such as
the Markov approximation and the rotating-wave approximation.
In principle, the derivation implies that 
the present model is justified only within  
a certain range of parameters.
However, the present setting without any additional constraints
is completely consistent with the laws of thermodynamics, and 
we can discuss the performance of the quantum heat engine.
The use of the jump operators projected 
onto the instantaneous eigenstates of the Hamiltonian 
is an important ingredient to find a thermodynamically consistent theory.
It is contrasted to the ``local'' master-equation approach in which
the jump operators are not projected to the eigenstate basis.
The local approach is shown to be inconsistent with 
thermodynamics~\cite{Levy14}.
Although some ideas to overcome this shortcoming have been 
discussed~\cite{Hewgill21,Chiara18}, 
here we use the ``global'' approach 
defined by Eq.~(\ref{dissipator}).

In principle, the present model has four types of 
the dissipator coupling, 
$\gamma_{\rm c}(\epsilon_{10})$, 
$\gamma_{\rm c}(\epsilon_{20})$, 
$\gamma_{\rm h}(\epsilon_{10})$, and 
$\gamma_{\rm h}(\epsilon_{20})$, where 
$\epsilon_{10}=\epsilon_1-\epsilon_0$
and $\epsilon_{20}=\epsilon_2-\epsilon_0$.
In previous works, 
only $\gamma_{\rm c}(\epsilon_{10})$ and $\gamma_{\rm h}(\epsilon_{20})$ 
were kept nonzero~\cite{Geva94,Geva96}.
As we mentioned above, 
$\gamma_{\alpha}(\epsilon)$ is a system-dependent function
and is obtained from the correlation function of a bath operator.
A typical form is represented by using a Lorentzian function.
In the following, we do not assume any functional form of 
$\gamma_{\alpha}(\epsilon)$ and 
consider three possible cases:
\begin{itemize}
\item[(i)] Resonant coupling 
\begin{eqnarray}
 \gamma_{\rm c}(\epsilon_{10})=\gamma_{\rm h}(\epsilon_{20})
 >\gamma_{\rm c}(\epsilon_{20})=\gamma_{\rm h}(\epsilon_{10})=0.
\end{eqnarray}
\item[(ii)] Intermediate coupling 
\begin{eqnarray}
 \gamma_{\rm c}(\epsilon_{10})=\gamma_{\rm h}(\epsilon_{20})
 >\gamma_{\rm c}(\epsilon_{20})=\gamma_{\rm h}(\epsilon_{10})>0.
\end{eqnarray}
\item[(iii)] Uniform coupling 
\begin{eqnarray}
 \gamma_{\rm c}(\epsilon_{10})=\gamma_{\rm h}(\epsilon_{20})
 =\gamma_{\rm c}(\epsilon_{20})=\gamma_{\rm h}(\epsilon_{10})>0.
\end{eqnarray}
\end{itemize}
The resonant-coupling case corresponds to the preceding works.

\subsection{Heat, work, and efficiency}

\begin{figure}[t]
\includegraphics[width=0.80\columnwidth]{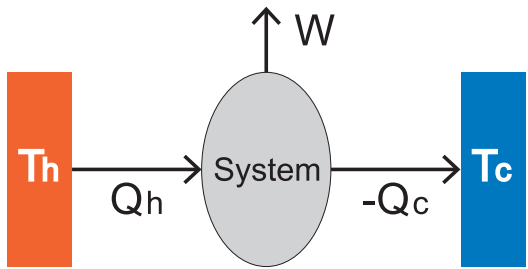}
\caption{An expected heat flow.
$W=Q_{\rm h}+Q_{\rm c}$, and 
the efficiency is given by Eq.~(\ref{efficiency}).}
\label{fig02}
\end{figure}

According to the standard scenario, we define the heat flux from 
the reservoirs to the system as
\begin{equation}
 \dot{Q}(t)={\rm Tr}\left[\partial_t\hat{\rho}(t)\hat{H}(t)\right]
 =\sum_\alpha{\rm Tr}\left[\hat{D}_\alpha[\hat{\rho}(t)]\hat{H}(t)\right].
 \label{heat}
\end{equation}
The last expression allows us to write 
$\dot{Q}(t)=\sum_\alpha\dot{Q}_\alpha(t)$, 
where $\dot{Q}_\alpha(t)$ represents 
the heat flux from each reservoir.
When the system is operated periodically,
the energy goes back to the original value after one period 
such that the work done by the system in the one period 
is given by
\begin{equation}
 W=\sum_\alpha Q_\alpha
 = \sum_\alpha \int_{T}^{T+T_0} dt\,\dot{Q}_\alpha(t),
\end{equation}
where $T_0=2\pi/\omega$. 
When the system acts as a heat engine,
the relations $Q_{\rm h}>0$, $Q_{\rm c}<0$, and $W>0$ hold, 
as we see in Fig.~\ref{fig02}, 
and the efficiency is defined as 
\begin{equation}
 \eta=\frac{W}{Q_{\rm h}}=1+\frac{Q_{\rm c}}{Q_{\rm h}}.
 \label{efficiency}
\end{equation}

The second law of thermodynamics is derived from the nonnegativity of
the entropy production~\cite{Alicki79}.
For the present model we obtain 
\begin{equation}
 -\sum_\alpha \beta_\alpha Q_\alpha \ge 0, 
\end{equation}
and as a result, the efficiency is bounded from above by 
the Carnot efficiency
\begin{equation}
 \eta^{\rm C}=1-\frac{T_{\rm c}}{T_{\rm h}}.
\end{equation}
The nonnegativity of the entropy production holds 
even in the present model~\cite{Funo19}, 
which confirms the previous result on the upper bound on 
the efficiency~\cite{Scovil59,Kosloff84,Geva94,Geva96,Boukobza06,Boukobza07}.

\section{Power and efficiency}
\label{sec:pandeff}

\subsection{Stationary solution}

We use the stationary solution of the GKLS equation in the present setting 
to evaluate the performance of the heat engine.
We note that the stationary solution means that 
the system settles down to a stable periodic behavior 
after transient evolutions during the first several periods.
The eigenstate decomposition of the jump operator 
$\langle \epsilon_n(t)|\hat{L}_{\alpha}|\epsilon_m(t)\rangle$
is shown to be time independent, which gives a simple stationary result.
We describe the details in Appendix~\ref{app-ss}.
Here we summarize the result.

The stationary solution of the GKLS equation was studied 
in the preceding works.
As we stressed above, the crucial difference is that the dissipator is 
represented by four types of dissipator couplings.
Correspondingly, the explicit form of the dissipator 
is parametrized by four types of coupling functions, 
\begin{eqnarray}
 && g_1=\gamma_{\rm c}(\epsilon_{10})\frac{1+\cos\theta}{2}
 +\gamma_{\rm h}(\epsilon_{10})\frac{1-\cos\theta}{2}, \label{g1}\\ 
 && g_2=\gamma_{\rm h}(\epsilon_{20})\frac{1+\cos\theta}{2}
 +\gamma_{\rm c}(\epsilon_{20})\frac{1-\cos\theta}{2}, \\
 && g_1^-=\gamma_{\rm c}(-\epsilon_{10})\frac{1+\cos\theta}{2}
 +\gamma_{\rm h}(-\epsilon_{10})\frac{1-\cos\theta}{2}, \\ 
 && g_2^-=\gamma_{\rm h}(-\epsilon_{20})\frac{1+\cos\theta}{2}
 +\gamma_{\rm c}(-\epsilon_{20})\frac{1-\cos\theta}{2}, \label{g2m}
\end{eqnarray}
where $\theta$ is defined by the relation
\begin{equation}
 \tan\theta=\frac{2\lambda}{\omega_2-\omega_1}.
 \label{theta}
\end{equation}
$\theta$ represents a rotation angle for the diagonalization of 
the Hamiltonian, as shown in Appendix \ref{app-eigen}.
In the resonant-coupling limit, relations 
$g_1^-=e^{-\beta_{\rm c}\epsilon_{10}}g_1$ and 
$g_2^-=e^{-\beta_{\rm h}\epsilon_{20}}g_2$ hold, and 
the dissipator is characterized by $g_1$ and $g_2$.
In the general case, no trivial relations hold 
between $g_1$, $g_2$, $g_1^-$, and $g_2^-$,
although we still have the detailed balance condition in Eq.~(\ref{dbc}).

We also introduce dimensionless nonnegative parameters 
\begin{eqnarray}
 && q_1 = \frac{\gamma_{\rm h}(\epsilon_{10})
 \frac{1-\cos\theta}{2}}{\gamma_{\rm c}(\epsilon_{10})
 \frac{1+\cos\theta}{2}+\gamma_{\rm h}(\epsilon_{10})
 \frac{1-\cos\theta}{2}}, \label{q1}\\
 && q_2 = \frac{\gamma_{\rm c}(\epsilon_{20})
 \frac{1-\cos\theta}{2}}{\gamma_{\rm h}(\epsilon_{20})
 \frac{1+\cos\theta}{2}+\gamma_{\rm c}(\epsilon_{20})
 \frac{1-\cos\theta}{2}}, \label{q2}
\end{eqnarray}
to characterize thermodynamic quantities in the following.
For the resonant-coupling case 
these quantities collapse to zero $q_1=q_2=0$.

At the stationary limit, 
the heat flux from each reservoir is given by  
\begin{eqnarray}
 && \dot{Q}_{\rm c}(t) \to \left[
 \frac{\epsilon_{20}}{\epsilon_{21}}q_2
 -\frac{\epsilon_{10} }{\epsilon_{21}}(1-q_1)
 \right]P - \rho_0P_0, \label{qc}\\
 && \dot{Q}_{\rm h}(t) \to  \left[
 \frac{\epsilon_{20}}{\epsilon_{21}}(1-q_2)
 -\frac{\epsilon_{10} }{\epsilon_{21} }q_1
 \right]P
 +\rho_0P_0, \label{qh}
\end{eqnarray}
where $\epsilon_{mn}=\epsilon_m-\epsilon_n$ and 
\begin{eqnarray}
 && P = 
 \epsilon_{21}\frac{\omega^2\sin^2\theta}{2G}
 \left(\frac{g_2^{-}}{g_2}-\frac{g_1^{-}}{g_1}\right)\frac{1}{Z}, \label{p} \\
 && Z= \left[1+\frac{\omega^2\sin^2\theta}{2G}
 \left(\frac{1}{g_1}+\frac{1}{g_2}\right)\right]
 \left(1+\frac{g_1^-}{g_1}+\frac{g_2^-}{g_2}\right)
 \nonumber\\
 &&  +\frac{\omega^2\sin^2\theta}{2G}\left(\frac{1}{g_1}-\frac{1}{g_2}\right)
 \left(\frac{g_2^{-}}{g_2}-\frac{g_1^{-}}{g_1}\right),
 \\
 && G = \frac{1}{2}(g_1+g_2)
 +\frac{\left(\epsilon_{21}-\omega\cos\theta\right)^2}{\frac{1}{2}(g_1+g_2)}, \\
 && \rho_0 = \frac{1}{1+\frac{g_1^{-}}{g_1}+\frac{g_2^{-}}{g_2}}
 \left[1-\left(\frac{1}{g_1}-\frac{1}{g_2}\right)
 \frac{P}{\epsilon_{21}}\right], \label{rho0} \\
 && P_0 = 
 \epsilon_{10}g_1q_1(1-q_1)
 \left(e^{-\beta_{\rm h}\epsilon_{10}}-e^{-\beta_{\rm c}\epsilon_{10}}\right) \nonumber\\
 && \qquad +\epsilon_{20}g_2q_2(1-q_2)
 \left(e^{-\beta_{\rm h}\epsilon_{20}}-e^{-\beta_{\rm c}\epsilon_{20}}\right).
\end{eqnarray}
All the quantities introduced above are independent of $t$.
$\rho_0$ and $G$ are always positive, irrespective of the choice of
the parameters.
We show below that $P$ represents the power of the heat engine 
and $\rho_0$ represents the ground-state component of the density operator, 
$\langle 0|\hat{\rho}(t)|0\rangle$.
The term $\rho_0 P_0$ represents a direct flow from the hot reservoir
to the cold reservoir, and  $P_0$ goes to zero when 
the temperature difference $\beta_{\rm c}-\beta_{\rm h}$ disappears.
Further details are discussed below.

\subsection{Power and efficiency}

Anticipating $Q_{\rm h}>0$, $Q_{\rm c}<0$, and $W>0$, 
we can obtain the explicit form of the work done by the system.
The power of the heat engine, 
which is defined by the work divided by the cycle period,
is given by
\begin{equation}
 \frac{W}{T_0}\to P \label{W}
\end{equation}
at the stationary limit.
$P$ is given in Eq.~(\ref{p}).

The corresponding efficiency is obtained as 
\begin{equation}
 \eta = \eta^{\rm SSD}\frac{\frac{1}{\eta^{\rm SSD}}P}{\frac{1}{\eta^{\rm SSD}}P
 +\rho_0P_0},
 \label{eta}
\end{equation}
where 
\begin{equation}
 \eta^{\rm SSD}=\frac{1}
 {1-q_2-\frac{\epsilon_{10}}{\epsilon_{20}}q_1}
 \left(1-\frac{\epsilon_{10}}{\epsilon_{20}}\right). \label{etaq}
\end{equation}
$\eta^{\rm SSD}$ is reminiscent of the Scovil--Schulz-DuBois 
efficiency~\cite{Scovil59}.
This expression is simplified when we consider the resonant-coupling limit 
$\gamma_{\rm c}(\epsilon_{20})=\gamma_{\rm h}(\epsilon_{10})=0$.
In this case, we find $q_1=0$, $q_2=0$, and $P_0=0$, 
and the efficiency coincides with the Scovil--Schulz-DuBois efficiency 
\begin{equation}
 \eta =\eta^{\rm SSD} \to  1-\frac{\epsilon_{10}}{\epsilon_{20}}. \label{etassd}
\end{equation}
This result is consistent with the preceding 
works~\cite{Kosloff84,Geva94,Geva96,Boukobza06,Boukobza07,Uzdin15,Singh20}.
The efficiency in the general case 
is dependent on various parameters 
such as the temperatures and the frequency 
and is bounded from above by Eq.~(\ref{etassd}).

\subsection{Heat engine conditions}

The above results in the present section are exact and 
hold irrespective of the choice of parameters.
When we require that the system works as a heat engine,
the relations $Q_{\rm h}>0$, $Q_{\rm c}<0$, and $P>0$ must hold. 

As we see from Eq.~(\ref{p}), $P>0$ holds when 
\begin{equation}
 \frac{g_1^-}{g_1}<\frac{g_2^-}{g_2}. \label{cond0}
\end{equation}
This condition is satisfied only when 
\begin{equation}
 \beta_{\rm c}\epsilon_{10}>\beta_{\rm h}\epsilon_{20}.  \label{cond2}
\end{equation}
This is a necessary condition in general 
and is the necessary and sufficient condition in the resonant-coupling case.
In the general case, Eq.~(\ref{cond0}) is rewritten as 
\begin{equation}
 \frac{q_1}{q_{10}}+\frac{q_2}{q_{20}} < 1,
 \label{cond1}
\end{equation}
where 
\begin{eqnarray}
 &&q_{10}=\frac{e^{-\beta_h \epsilon_{20}} - e^{-\beta_c \epsilon_{10}}}
 {e^{-\beta_h \epsilon_{10}} - e^{-\beta_c \epsilon_{10}}}, \\
 && q_{20}=\frac{e^{-\beta_h \epsilon_{20}} - e^{-\beta_c \epsilon_{10}}}
 {e^{-\beta_h \epsilon_{20}} - e^{-\beta_c \epsilon_{20}}}.
 \label{q10q20}
\end{eqnarray}
The coupling-constant parameters in the dissipator are taken 
so that Eq.~(\ref{cond1}) is satisfied.
We note that $\eta^{\rm SSD}$ in Eq.~(\ref{etaq}) becomes positive in that case.
The condition in Eq.~(\ref{cond2}) determines possible values of parameters 
in the Hamiltonian for a given $\eta^{\rm C}$ as 
\begin{eqnarray}
 &&\frac{\omega_{20}}{\omega_{10}}< \frac{1}{1-\eta^{\rm C}}, \label{cond21}\\
 &&\lambda^2<\frac{\left[\frac{\omega_{20}}{\omega_{10}}-(1-\eta^{\rm C})\right]
 \left[1-(1-\eta^{\rm C})\frac{\omega_{20}}{\omega_{10}}\right]}
 {(2-\eta^{\rm C})^2}.\label{cond22}
\end{eqnarray}

Equation (\ref{cond2}) shows that 
the power becomes negative when the temperature difference is too small.
In that case, the system does not work as a heat engine, and
we can observe a heat flow from the low-temperature reservoir 
to the high-temperature one as $Q_{\rm h}<0$ and $Q_{\rm c}>0$.
This behavior implies that the present system cannot be understood from
the standard linear-response theory in which the heat flow 
arises due to the temperature difference.
We discuss the origin of this quantum nature in the next section.

\subsection{Plot of the results}

\begin{figure}[t]
\includegraphics[width=0.60\columnwidth]{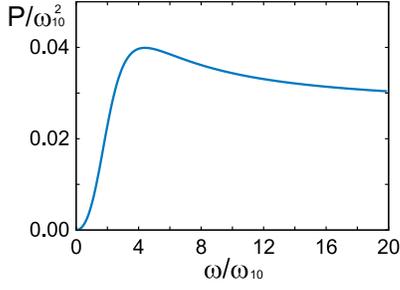}
\caption{Typical behavior of the power $P$ 
as a function of the frequency $\omega$.
$P$ has a peak at the frequency given in Eq.~(\ref{omega}) and 
approaches a positive finite value at the limit $\omega\to\infty$.
We set $\beta_{\rm c}\omega_{10}=5.0$, 
$\beta_{\rm h}\omega_{10}=1.0$, 
$\frac{\gamma_{\rm c}(\epsilon_{10})}{\omega_{10}}
=\frac{\gamma_{\rm h}(\epsilon_{20})}{\omega_{10}}=2.0$, 
$\gamma_{\rm c}(\epsilon_{20})=\gamma_{\rm h}(\epsilon_{10})=0$, 
$\frac{\lambda}{\omega_{10}}=0.5$, and
$\frac{\omega_{20}}{\omega_{10}}=2.5$.
}
\label{fig03}
\end{figure}

At small frequency values, the power $P$ is proportional to $\omega^2$.
The efficiency $\eta$ is also proportional to $\omega^2$ 
provided $P_0>0$, as we see from Eq.~(\ref{eta}).
We see that the frequency-independent result in Eq.~(\ref{etassd})
is specific to the resonant coupling and is unusual.

$P$ as a function of $\omega$
has a Fano resonant form 
and is plotted in Fig~\ref{fig03}.
It is maximized at 
\begin{equation}
 \omega=\frac{\epsilon_{21}^2+\frac{1}{4}(g_1+g_2)^2}{\omega_2-\omega_1}.
 \label{omega}
\end{equation}
Interestingly, the efficiency $\eta$ 
is also maximized at this frequency.
The magnitude of the optimal frequency is basically determined by 
the energy gap $\epsilon_{21}$.
The dissipation effect enhances the resonant frequency slightly.

\begin{figure}[t]
\includegraphics[width=1.0\columnwidth]{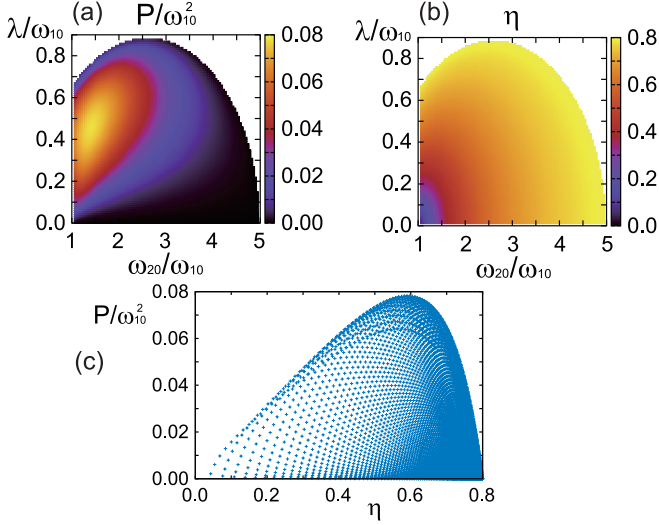}
\caption{
The power $P$ and the efficiency $\eta$ at the resonant coupling 
$\frac{\gamma_{\rm c}(\epsilon_{10})}{\omega_{10}}
=\frac{\gamma_{\rm h}(\epsilon_{20})}{\omega_{10}}=2.0$ and 
$\gamma_{\rm c}(\epsilon_{20})=\gamma_{\rm h}(\epsilon_{10})=0$.
We set $\beta_{\rm c}\omega_{10}=5.0$ and $\beta_{\rm h}\omega_{10}=1.0$.
The frequency is chosen as in Eq.~(\ref{omega}).
(a) $P$ as a function of the parameters in the Hamiltonian.
(b) $\eta$ as a function of the parameters in the Hamiltonian.
(c) Distributions of $(\eta, P)$.
We note that the Carnot efficiency is given by $\eta^{\rm C}=0.8$ 
in the present choice of parameters.
}
\label{fig04}
\end{figure}
\begin{figure}[t]
\includegraphics[width=1.0\columnwidth]{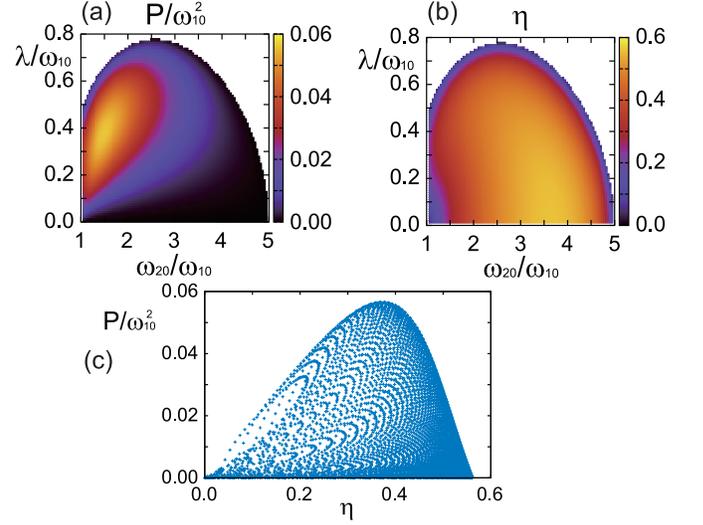}
\caption{
The power $P$ and the efficiency $\eta$ at the intermediate coupling 
$\frac{\gamma_{\rm c}(\epsilon_{10})}{\omega_{10}}
=\frac{\gamma_{\rm h}(\epsilon_{20})}{\omega_{10}}=2.0$ and 
$\frac{\gamma_{\rm c}(\epsilon_{20})}{\omega_{10}}
=\frac{\gamma_{\rm h}(\epsilon_{10})}{\omega_{10}}=0.5$.
See the caption of Fig.~\ref{fig04} for other remarks.
}
\label{fig05}
\end{figure}
\begin{figure}[t]
\includegraphics[width=1.0\columnwidth]{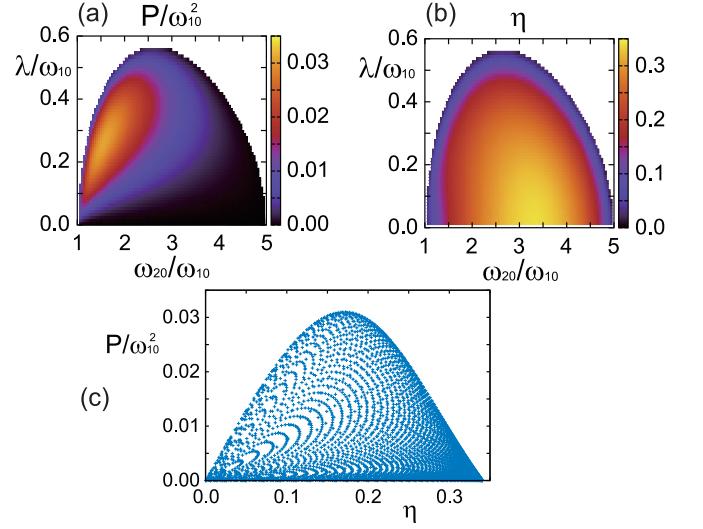}
\caption{
The power $P$ and the efficiency $\eta$ at the uniform coupling 
$\frac{\gamma_{\rm c}(\epsilon_{10})}{\omega_{10}}
=\frac{\gamma_{\rm h}(\epsilon_{20})}{\omega_{10}} 
=\frac{\gamma_{\rm c}(\epsilon_{20})}{\omega_{10}}
=\frac{\gamma_{\rm h}(\epsilon_{10})}{\omega_{10}}=2.0$.
See the caption of Fig.~\ref{fig04} for other remarks.
}
\label{fig06}
\end{figure}

We plot the power $P$ and the efficiency $\eta$
for the optimal frequency $\omega$ in Eq.~(\ref{omega})
in Fig.~\ref{fig04} (resonant coupling), 
Fig.~\ref{fig05} (intermediate coupling), 
and Fig.~\ref{fig06} (uniform coupling). 
We set $\beta_{\rm c}\omega_{10}=5.0$ and $\beta_{\rm h}\omega_{10}=1.0$,
which gives the Carnot efficiency $\eta^{\rm C}=0.8$.
$P$ and $\eta$ are plotted as functions of $\omega_{20}$ and $\lambda$
under the heat-engine conditions 
in Eqs.~(\ref{cond1}), (\ref{cond21}), and (\ref{cond22}).

We observe that the power is maximized 
at small $\omega_{20}/\omega_{10}$ ($>1$) and at a moderate value of $\lambda$.
Although the possible parameter range for a heat engine 
is dependent on the dissipator couplings as well as the temperatures,
the contour map is basically insensitive to the parameters.

In contrast to the power, 
the efficiency exhibits a stronger dependence on the dissipator coupling. 
As shown in Fig.~\ref{fig04}(b) for the resonant coupling, 
the efficiency attains its maximum 
at the boundary where the power goes to zero.
This behavior is totally reversed 
in comparison to the other dissipator coupling cases
in Figs.~\ref{fig05}(b) and \ref{fig06}(b), 
where the efficiency is minimized at the boundary.
We also find that the efficiency and the power 
are reduced when we move away from the resonant coupling.

We plot the distribution ($\eta,P$)
in Figs.~\ref{fig04}(c), \ref{fig05}(c), and \ref{fig06}(c).
We find that the envelope curve of the distributions has 
a tendency towards being symmetric around the efficiency 
at maximum power as we approach the uniform coupling. 
We note that the efficiency at maximum power cannot be understood from 
Curzon--Ahlborn efficiency even in the linear-response 
regime~\cite{Curzon75,Vandenbroeck05}.
As we mentioned above, the present system 
does not exhibit a heat-engine behavior at the linear response regime.

Summarizing the present result, we find that 
the performance of the present model as a heat engine 
worsens when we move away from the resonant coupling.
In the next section, we discuss the origin of this behavior.

\section{Decomposition of heat}
\label{sec:decomp}

Compared to the resonant coupling, we see that 
the decreasing of the efficiency in extended couplings
is understood from the presence of $\rho_0P_0$ in Eq.~(\ref{eta}).
As we also see in Eqs.~(\ref{qc}) and (\ref{qh}), 
it represents a direct flow from the hot reservoir to the cold reservoir,
which clearly reduces the performance of the heat engine.
$P_0$ goes to zero when the temperature difference is zero and 
remains positive even at the zero-field limit $\lambda\to 0$.
This implies that the direct flow has a classical interpretation.

The quantum nature of the system can be understood by 
realizing that the basis of the density operator is
different from that of the Hamiltonian.
The density operator is diagonalized as 
\begin{equation}
 \hat{\rho}(t)=\sum_n p_n|\rho_n(t)\rangle\langle\rho_n(t)|.
 \label{rho}
\end{equation}
The eigenvalues of the density operator obey 
the master-equation-like relation
\begin{equation}
 \partial_tp_n= \sum_\alpha\langle\rho_n(t)|\hat{D}_\alpha[\hat{\rho}(t)]
 |\rho_n(t)\rangle, \label{master}
\end{equation}
and the basis of the density operator obeys the unitary time evolution 
\begin{equation}
 i\partial_t|\rho_n(t)\rangle=\hat{\xi}(t)|\rho_n(t)\rangle,
\end{equation}
where $\hat{\xi}(t)$ represents a generator of 
the time evolution~\cite{Funo19}.
The explicit forms of the eigenvalues and the eigenstates of 
the density operator are given in Appendix~\ref{app-rho}.

Accordingly, the heat flux in Eq.~(\ref{heat}) is decomposed into 
the diagonal and nondiagonal parts as 
$\dot{Q}_{\alpha}(t)=\dot{Q}_{\alpha}^{\rm d}(t)+\dot{Q}_{\alpha}^{\rm nd}(t)$, where 
\begin{equation}
 \dot{Q}_{\alpha}^{\rm d}(t)=
 \sum_n\langle\rho_n(t)|\hat{D}_\alpha[\hat{\rho}(t)]|\rho_n(t)\rangle
 \langle\rho_n(t)|\hat{H}(t)|\rho_n(t)\rangle
\end{equation}
and $\dot{Q}_{\alpha}^{\rm nd}(t)$ is defined by the residual contribution
of $\dot{Q}_{\alpha}(t)$.
The diagonal part is related to the population dynamics 
in Eq.~(\ref{master}), 
and the nondiagonal part is related to the coherent dynamics.
In the present setting, the eigenvalue $p_n$ is independent of $t$.
As a result, we obtain 
\begin{equation}
 \dot{Q}_{\rm h}^{\rm d}+\dot{Q}_{\rm c}^{\rm d}=0. \label{qd}
\end{equation}
This relation shows that the diagonal part of the heat flux 
just goes through the system as  
$\dot{Q}^{\rm d}_{\rm h}=-\dot{Q}^{\rm d}_{\rm c}$
and does not contribute to the work.
This direct flow reduces the efficiency of the heat engine. 

To quantify the diagonal and nondiagonal contributions, 
we decompose the efficiency as 
\begin{equation}
 \eta = \eta^{\rm d}\frac{Q_{\rm h}^{\rm d}}{Q_{\rm h}}
 +\eta^{\rm nd}\frac{Q_{\rm h}^{\rm nd}}{Q_{\rm h}}, \label{etadnd}
\end{equation}
where 
\begin{eqnarray}
 &&\eta^{\rm d}=1+\frac{Q_{\rm c}^{\rm d}}{Q_{\rm h}^{\rm d}}, \\
 &&\eta^{\rm nd}=1+\frac{Q_{\rm c}^{\rm nd}}{Q_{\rm h}^{\rm nd}}.
\end{eqnarray}
$\eta^{\rm d}$ denotes the contribution from the diagonal part
of the heat flux, and $\eta^{\rm nd}$ denotes 
the contribution from the nondiagonal part.
Equation (\ref{qd}) shows that $\eta^{\rm d}=0$ in the present case.
The efficiency is given by 
\begin{equation}
 \eta = \eta^{\rm nd}
 \frac{\frac{1}{\eta^{\rm nd}}P}{\frac{1}{\eta^{\rm SSD}}P+\rho_0P_0}, \label{eta2}
\end{equation}
and the explicit form of $\eta^{\rm nd}$ is 
\begin{eqnarray}
 \frac{1}{\eta^{\rm nd}} &=& 
 \frac{g_1q_1G_1+g_2(1-q_2)G_2}{g_1G_1+g_2G_2}
 \nonumber\\
 && +\frac{-g_1(1-q_1)\frac{q_1}{q_{10}}+g_2(1-q_2)\frac{q_2}{q_{20}}}
 {1-\frac{q_1}{q_{10}}-\frac{q_2}{q_{20}}}
 \frac{G_1+G_2}{g_1G_1+g_2G_2}, \nonumber\\
\end{eqnarray}
where $G_1=G+\frac{4\lambda^2\omega^2}{g_1\epsilon_{21}^2}$ and 
$G_2=G+\frac{4\lambda^2\omega^2}{g_2\epsilon_{21}^2}$.
From Eq.~(\ref{eta2}), we can identify the diagonal and 
nondiagonal heat flows, respectively, as
\begin{eqnarray}
 && \frac{Q_{\rm h}^{\rm d}}{T_0}=-\frac{Q_{\rm c}^{\rm d}}{T_0}
 =\left(\frac{1}{\eta^{\rm SSD}}-\frac{1}{\eta^{\rm nd}}\right)P+\rho_0P_0, \\
 && \frac{Q_{\rm h}^{\rm nd}}{T_0}=\frac{1}{\eta^{\rm nd}}P.
\end{eqnarray}

\begin{figure}[t]
\includegraphics[width=1.0\columnwidth]{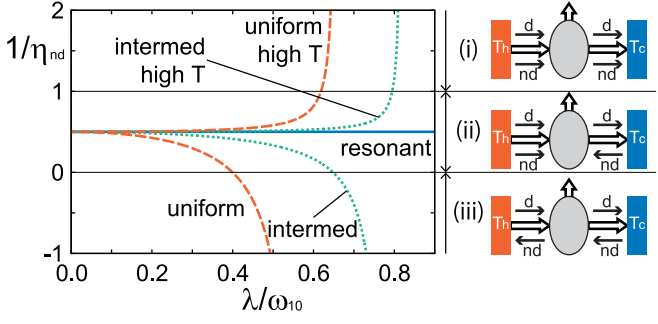}
\caption{
Left: $1/\eta^{\rm nd}$ as a function of $\lambda$ 
for $\omega_{20}/\omega_{10}=2.6$.
We take 
$\beta_{\rm c}\omega_{10}=5.0$, $\beta_{\rm h}\omega_{10}=1.0$ 
for plots of decreasing functions
and 
$\beta_{\rm c}\omega_{10}=1.0$, $\beta_{\rm h}\omega_{10}=0.2$ 
for plots of increasing functions with high $T$.
We use the parametrizations of the dissipator couplings
in Fig.~\ref{fig04} (for ``resonant''), 
Fig.~\ref{fig05} (for ``intermed"), and 
Fig.~\ref{fig06} (for ``uniform'').
The frequency is chosen as in Eq.~(\ref{omega}).
Right: Sketch of heat flows.
Each arrow denotes the direction of the flow.
``d'' denotes the diagonal part, and ``nd'' denotes the nondiagonal part.
}
\label{fig07}
\end{figure}
\begin{figure}[t]
\includegraphics[width=1.0\columnwidth]{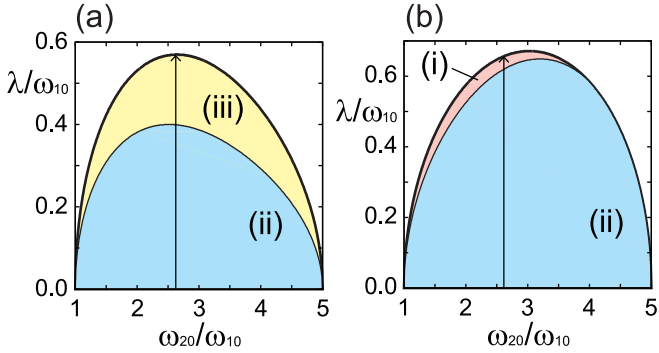}
\caption{
Three possible patterns of heat flow drawn in Fig.~\ref{fig07}
in the uniform-coupling case.
(a) $\beta_{\rm c}\omega_{10}=5.0$ and $\beta_{\rm h}\omega_{10}=1.0$.
(b) $\beta_{\rm c}\omega_{10}=1.0$ and $\beta_{\rm h}\omega_{10}=0.2$. 
The arrows denote red dashed lines in Fig.~\ref{fig07}.
}
\label{fig08}
\end{figure}

We plot $1/\eta_{\rm nd}$ as a function of $\lambda$ in Fig.~\ref{fig07}.
From the value of $1/\eta_{\rm nd}$, 
we can understand the pattern of the heat flow.
When $\eta_{\rm nd}>1$, both $Q_{\rm c}^{\rm nd}$ and $Q_{\rm h}^{\rm nd}$ are positive,
and we can obtain the ideal heat flow as a heat engine.
The case $\eta_{\rm nd}>1$ is represented by pattern (ii) in Fig.~\ref{fig07}.
$\eta_{\rm nd}$ has a purely quantum-mechanical origin and 
can exceed the Carnot efficiency.
However, we cannot neglect 
the diagonal contribution, which reduces the total efficiency.
As a result, even in the quantum system, 
the efficiency in Eq.~(\ref{eta2}) is bounded from above 
by the Carnot efficiency.

In the resonant-coupling case, $\eta_{\rm nd}$ is always larger than unity,
which leads to a high-performance result in Fig.~\ref{fig04}.
The situation is significantly changed when we move away from 
the resonant-coupling case.
As we increase $\lambda$, the nondiagonal heat flow changes, 
and we observe the reduction of the efficiency as a result.
Dependent on the parameters in the equation, 
we can observe pattern (i) and pattern (iii) in Fig.~\ref{fig07} 
within the heat-engine domain.

In Fig.~\ref{fig08}, we plot the patterns of the heat flow 
in the uniform-coupling case.
We can find patterns (i) and (iii) around the boundary 
where the efficiency becomes small.
This behavior is consistent with that in Fig.~\ref{fig06}.

The decomposition of the diagonal part and the nondiagonal part
has been discussed in some works~\cite{Baumgratz14,Santos19}.
It is a difficult problem to observe each one as an independent contribution.
However, the nondiagonal part goes to zero at $\lambda\omega\to 0$.
The diagonal part is insensitive to the parameter and 
can be extracted in the weak-field regime.

\section{Thermodynamic uncertainty relation}
\label{sec:tur}

As a final subject to study, 
we examine the TUR~\cite{Barato15,Gingrich16}.
The standard form of the TUR is represented as 
\begin{equation}
 \langle\dot{\sigma}\rangle\frac{{\rm var}\,P}{P^2}\ge 2, \label{tur}
\end{equation}
where $\langle\dot{\sigma}\rangle$ is the entropy production rate 
averaged over one cycle and 
${\rm var}\,P$ is the variance of the power.
The variance is bounded from below, and 
the bound is determined by the entropy production.
Although this relation was shown in a broad range of classical systems, 
the relation does not necessarily hold, and the violation can be found 
especially in quantum systems. 
The quantum TUR is modified by a different bound~\cite{Hasegawa20}.
For the present three-level system, 
the violation of the standard TUR was shown in \cite{Kalaee21,Menczel21}.
Here we examine how this result is affected by the modification of 
the dissipator.

The entropy production rate at each $t$ is given by 
$\dot{\sigma}(t)=-{\rm Tr}\,\partial_t\hat{\rho}(t)\ln\hat{\rho}(t)
-\sum_\alpha\beta_{\alpha}\dot{Q}_\alpha(t)$.
The first term comes from the von Neumann entropy of the system 
and goes to zero when we take the average over one period.
The average of the entropy production rate 
 $\langle\dot{\sigma}\rangle 
 =\lim_{T\to\infty}\int_{T}^{T+T_0}dt\,\dot{\sigma}(t)/T_0$
is calculated as 
\begin{equation}
 \langle\dot{\sigma}\rangle = (\beta_{\rm c}-\beta_{\rm h})
 \left[\left(\frac{1}{\eta^{\rm SSD}}-\frac{1}{\eta^{\rm C}}\right)P
 +\rho_0 P_0\right]. \label{ep}
\end{equation}

The variance of the power ${\rm var}\,P$ is calculated 
in Appendix \ref{app-tur}.
It is decomposed as 
${\rm var}P = ({\rm var}P)_1+({\rm var}P)_2
+({\rm var}P)_3+({\rm var}P)_4$, and each part is respectively given as 
\begin{eqnarray}
 ({\rm var}P)_1=\epsilon_{21}
 \frac{\frac{g_1^-}{g_1}+\frac{g_2^-}{g_2}}{\frac{g_2^-}{g_2}
 -\frac{g_1^-}{g_1}} P,
\label{varp1}
\end{eqnarray}
\begin{eqnarray}
 ({\rm var}P)_2
 &=&-\Biggl[
 \frac{
 \frac{1}{g_1}+\frac{1}{g_2}
 +\left(\frac{1}{g_1}-\frac{1}{g_2}\right)
 \left(\frac{g_2^-}{g_2}-\frac{g_1^-}{g_1}\right)
 }{1+\frac{g_1^-}{g_1}+\frac{g_2^-}{g_2}} 
 \nonumber\\ && 
 +\frac{1}{g_1}+\frac{1}{g_2}-\frac{4}{g_1+g_2}+\frac{4}{G}
 \Biggr]P^2,
\end{eqnarray}
\begin{eqnarray}
 ({\rm var}P)_3 &=& \Biggl[
 \left(\frac{1}{g_1}-\frac{1}{g_2}\right)^2
 \nonumber\\ && 
 +\left(\frac{1}{g_1}+\frac{1}{g_2}\right)
 \left(\frac{1}{g_1}+\frac{1}{g_2}-\frac{4}{g_1+g_2}+\frac{4}{G}\right)
 \nonumber\\ && 
 \times\left(1+\frac{g_1^-}{g_1}+\frac{g_2^-}{g_2}\right)
 \Biggr]\frac{P^3}{\epsilon_{21}\left(\frac{g_2^-}{g_2}
 -\frac{g_1^-}{g_1}\right)},
\end{eqnarray}
\begin{eqnarray}
 ({\rm var}P)_4 &=& \Biggl[
 \left(\frac{1}{g_1}+\frac{1}{g_2}\right)
 \left(1+\frac{1}{1+\frac{g_1^-}{g_1}+\frac{g_2^-}{g_2}} \right)
 \nonumber\\ && 
 +\left(\frac{1}{g_1}-\frac{1}{g_2}\right)
 \left(1+\frac{\frac{g_2^-}{g_2}-\frac{g_1^-}{g_1}}{1+\frac{g_1^-}{g_1}
 +\frac{g_2^-}{g_2}} \right)
 \nonumber\\ && 
 +\frac{1}{g_1}+\frac{1}{g_2}-\frac{4}{g_1+g_2}+\frac{4}{G}
 \Biggr]\frac{\left(\frac{1}{g_1}-\frac{1}{g_2}\right)P^3}{\epsilon_{21}}.
 \nonumber\\
\end{eqnarray}
We note that $({\rm var}\,P)_1$ and $({\rm var}\,P)_3$ are nonnegative.
We also see that $({\rm var}\,P)_2$ is negative, 
and $({\rm var}\,P)_4$ is zero at $g_1=g_2$.

\begin{figure}[t]
\includegraphics[width=1.0\columnwidth]{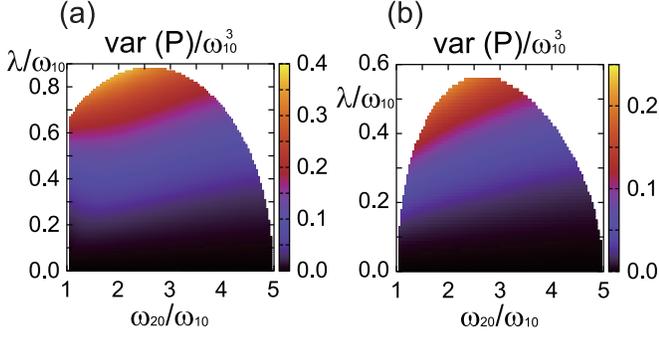}
\caption{
The variance of the power.
(a) Resonant-coupling case in Fig.~\ref{fig04}.
(b) Uniform-coupling case in Fig.~\ref{fig06}.
}
\label{fig09}
\end{figure}
\begin{figure}[t]
\includegraphics[width=1.0\columnwidth]{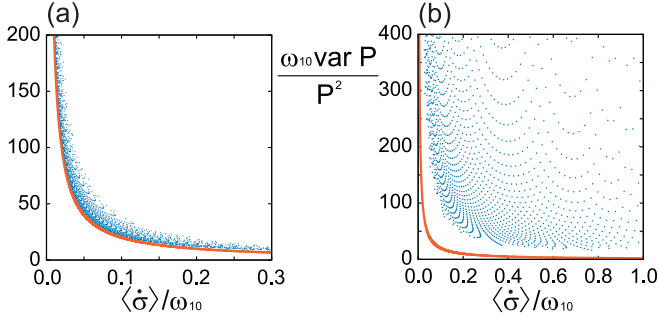}
\caption{The thermodynamic uncertainty relation.
(a) Resonant-coupling case in Fig.~\ref{fig04}.
(b) Uniform-coupling case in Fig.~\ref{fig06}.
The red lines denote the bound 
$\langle\dot{\sigma}\rangle{\rm var}(P)/P^2=2$.
}
\label{fig10}
\end{figure}
\begin{figure}[t]
\includegraphics[width=1.0\columnwidth]{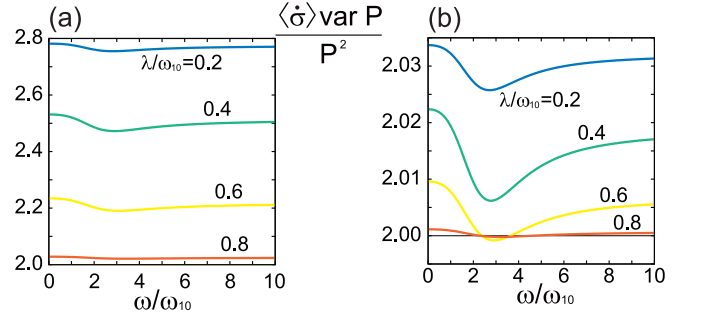}
\caption{
The frequency dependence of the thermodynamic uncertainty product
$\langle\dot{\sigma}\rangle{\rm var}(P)/P^2$ 
in the resonant-coupling case with $\omega_{20}/\omega_{10}=2.6$.
(a) $(\beta_{\rm c}\omega_{10}, \beta_{\rm h}\omega_{10})=(5.0,1.0)$.
(b) $(\beta_{\rm c}\omega_{10}, \beta_{\rm h}\omega_{10})=(1.0,0.2)$.
}
\label{fig11}
\end{figure}
\begin{figure}[t]
\includegraphics[width=1.0\columnwidth]{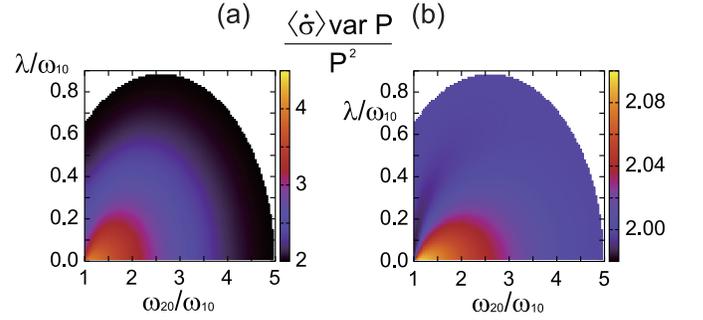}
\caption{
The thermodynamic uncertainty product
$\langle \dot{\sigma}\rangle{\rm var}(P)/P^2$ 
in the resonant-coupling case.
(a) $(\beta_{\rm c}\omega_{10}, \beta_{\rm h}\omega_{10})=(5.0,1.0)$.
(b) $(\beta_{\rm c}\omega_{10}, \beta_{\rm h}\omega_{10})=(1.0,0.2)$.
See the caption of Fig.~\ref{fig04} for the choice of the other parameters.
}
\label{fig12}
\end{figure}
\begin{figure}[t]
\includegraphics[width=1.0\columnwidth]{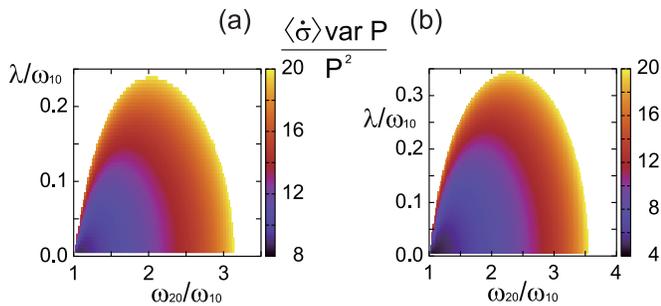}
\caption{
The thermodynamic uncertainty product
$\langle\dot{\sigma}\rangle{\rm var}(P)/P^2$ 
in the uniform-coupling case.
We plot the region where 
$\langle\dot{\sigma}\rangle{\rm var}(P)/P^2\le 20$.
(a) $(\beta_{\rm c}\omega_{10}, \beta_{\rm h}\omega_{10})=(5.0,1.0)$.
(b) $(\beta_{\rm c}\omega_{10}, \beta_{\rm h}\omega_{10})=(1.0,0.2)$.
See the caption of Fig.~\ref{fig06} for the choice of the other parameters.
}
\label{fig13}
\end{figure}

We plot ${\rm var}\,P$ in Fig.~\ref{fig09}.
The variance is basically 
an increasing function of $\lambda/\omega_{10}$ 
and is insensitive to $\omega_{20}/\omega_{10}$.
This result holds irrespective of the choice of the dissipator coupling.
The corresponding behavior of the TUR is shown in Fig.~\ref{fig10}.
As we see Fig.~\ref{fig10}, the TUR bound is strictly satisfied 
with the present choice of parameters.
The bound is tight in the case of the resonant coupling and 
is loosened as we move away from the resonant coupling.

All of the results in Figs.~\ref{fig09} and \ref{fig10} 
satisfy the standard TUR in Eq.~(\ref{tur}).
The optimal frequency in Eq.~(\ref{omega}) is used there, 
and the result is changed by considering the frequency dependence
in Fig.~\ref{fig11}.
We observe a violation of the standard TUR.
We show the contour maps of the uncertainty product 
in Figs.~\ref{fig12} and \ref{fig13}.
The violation 
occurs in a tiny range of parameters in the resonant-coupling case.

These numerical results can be understood from the analytical expression.
Basically, the bound comes from Eq.~(\ref{varp1}).
In the resonant-coupling case, we find 
\begin{equation}
 \langle\dot{\sigma}\rangle\frac{({\rm var}\,P)_1}{P^2}
 =\frac{\beta_{\rm c}\epsilon_{10}-\beta_{\rm h}\epsilon_{20}}
 {\tanh\frac{\beta_{\rm c}\epsilon_{10}-\beta_{\rm h}\epsilon_{20}}{2}} \ge 2.
\end{equation}
We numerically find that the other parts are very small, 
at least with the present choice of parameters.
The small violation of the TUR in Fig.~\ref{fig11} is understood from 
a negative contribution of $({\rm var}\,P)_2$.
We note that
the contributions $({\rm var}\,P)_1$ and $({\rm var}\,P)_2$ 
are similar to the result in Ref.~\cite{Kalaee21}, where 
the variance takes the form 
\begin{equation}
 {\rm var}\,P= AP-BP^2.
\end{equation}
$A$ and $B$ are positive, and the second term leads to a violation
of the TUR in a certain range of parameters.
We note that the local approach is used in Ref.~\cite{Kalaee21} 
and the form of the dissipator is different from the present model.
Our model includes additional contributions, 
$({\rm var}\,P)_3$ and $({\rm var}\,P)_4$,
but we numerically find that these contributions are small and
do not play any significant role.

In the case of the other dissipator coupling, 
we can understand the loose bound 
from the expression of the entropy production rate. 
The last term in Eq.~(\ref{ep}) includes $P_0$, 
which makes the uncertainty product
$\langle\dot{\sigma}\rangle{\rm var}(P)/P^2$ very large.

\section{Conclusions}
\label{sec:conc}

We have presented a detailed thermodynamic analysis 
of a continuous quantum heat engine based on the global form of 
the quantum master equation.  
We found that the performance of the heat engine 
is strongly dependent on the form of the dissipator. 
The quantum coherence does not necessarily 
enhance the efficiency of the heat engine.

The quantum coherent heat flow cannot be understood from the laws
of thermodynamics.
It produces a nontrivial heat flow even in the linear-response regime.
Although the coherent flow has the potential ability 
to enhance the efficiency of the heat engine, 
the efficiency is affected by the population heat flow, 
which prevents the system from violating the second law of thermodynamics. 
In the present model, the population heat flow does not contribute 
to the power of the heat engine but is required to keep 
the system within the heat-engine operation regime.

The decomposition of the efficiency in Eq.~(\ref{etadnd}) allowed us to 
discuss the interplay between the population and coherent parts.
We compared $\eta^{\rm d}$ and $\eta^{\rm nd}$ to find which part 
is important for the system to work as a heat engine.
In the present setting, we found $\eta^{\rm d}=0$.
Microscopically, this property is due to 
the time independence of the eigenvalues of the Hamiltonian.
When we consider a more general form of the Hamiltonian, 
the eigenvalue is generally dependent on $t$, and the population 
heat flow contributes to the power generation.
It will be an interesting problem to study the performance of the heat
engine and the interplay between the population and coherent parts
in that case.

\section*{Acknowledgements}

The authors are grateful to Yuki Izumida, Yusuke Nishida, 
Yasuhiro Tokura, and Yasuhiro Utsumi for useful discussions and comments.
P.B.-P. acknowledges the Japanese Government (MEXT) scholarship 
for undergraduate students for financial support.
K.T. was supported by JSPS KAKENHI Grants No. JP20K03781 and No. JP20H01827. 


\appendix
\section{Diagonalization of the Hamiltonian}
\label{app-eigen}

The Hamiltonian in Eq.~(\ref{hamiltonian}) is easily diagonalized 
as $\hat{H}(t)=\sum_n\epsilon_n|\epsilon_n(t)\rangle\langle\epsilon_n(t)|$.
The eigenvalues are independent of $t$ and are given by
\begin{eqnarray}
 && \epsilon_0=\omega_0, \\
 && \epsilon_1=\frac{\omega_0+\omega_1}{2}
 -\sqrt{\left(\frac{\omega_2-\omega_1}{2}\right)^2+\lambda^2}, \\
 && \epsilon_2=\frac{\omega_0+\omega_1}{2}
 +\sqrt{\left(\frac{\omega_2-\omega_1}{2}\right)^2+\lambda^2}.
\end{eqnarray}
$\epsilon_2$ is the largest eigenvalue under the condition
$\omega_0<\omega_1<\omega_2$.
The additional relation $\epsilon_0<\epsilon_1$ holds when 
we assume $\lambda^2< (\omega_1-\omega_0)(\omega_2-\omega_0)$.

The corresponding eigenstates are given by
\begin{eqnarray}
 && |\epsilon_0(t)\rangle = |0\rangle, \\
 && |\epsilon_1(t)\rangle = |1\rangle\cos\frac{\theta}{2}
 -|2\rangle e^{-i\omega t}\sin\frac{\theta}{2}, \\
 && |\epsilon_2(t)\rangle = |1\rangle e^{i\omega t}\sin\frac{\theta}{2}
 +|2\rangle\cos\frac{\theta}{2},
\end{eqnarray}
where $\theta$ is defined in Eq.~(\ref{theta}).

\section{Stationary solution of the GKLS equation}
\label{app-ss}

We obtain the stationary solution of the GKLS equation in Eq.~(\ref{gkls}).
Multiplying $\langle \epsilon_m(t)|$ from the left
and $|\epsilon_n(t)\rangle$ from the right in Eq.~(\ref{gkls}),
we obtain the expression
\begin{eqnarray}
 && \partial_t\langle \epsilon_m(t)|\hat{\rho}(t)|\epsilon_n(t)\rangle 
 \nonumber\\
 &=& - i (\epsilon_m-\epsilon_n)
 \langle\epsilon_m(t)|\hat{\rho}(t)|\epsilon_n(t)\rangle 
 \nonumber\\ && 
 +\sum_\alpha\langle\epsilon_m(t)|\hat{D}_\alpha[\hat{\rho}(t)]|\epsilon_n(t)
 \rangle \nonumber\\ && 
 +\langle\partial_t\epsilon_m(t)|\hat{\rho}(t)|\epsilon_n(t)\rangle 
 +\langle \epsilon_m(t)|\hat{\rho}(t)|\partial_t\epsilon_n(t)\rangle.
\end{eqnarray}
Due to the property $|\epsilon_0(t)\rangle=|0\rangle$ 
and our choice of the jump operators in Eqs.~(\ref{lc}) and (\ref{lh}),
$\langle 0|\hat{\rho}(t)|\epsilon_1(t)\rangle$ and
$\langle 0|\hat{\rho}(t)|\epsilon_2(t)\rangle$, their conjugates 
do not couple to the other components and decay exponentially 
as a function of $t$.
We write the off-diagonal component 
\begin{equation}
 \langle\epsilon_1(t)|\hat{\rho}(t)|\epsilon_2(t)\rangle
 =e^{i\omega t}\left(\Delta_1(t)+i\Delta_2(t)\right),
\end{equation}
where $\Delta_1$ and $\Delta_2$ are real, to obtain
\begin{eqnarray}
 && \partial_t\langle 0|\hat{\rho}|0\rangle =
 g_1\langle\epsilon_1|\hat{\rho}|\epsilon_1\rangle 
 +g_2\langle\epsilon_2|\hat{\rho}|\epsilon_2\rangle 
 -(g_1^-+g_2^-)\langle 0|\hat{\rho}|0\rangle, \nonumber\\ \\
 && \partial_t\langle \epsilon_1|\hat{\rho}|\epsilon_1\rangle =
 -g_1\langle\epsilon_1|\hat{\rho}|\epsilon_1\rangle 
 +g_1^-\langle 0|\hat{\rho}|0\rangle 
 -\omega \Delta_2\sin\theta , \\
 && \partial_t\langle \epsilon_2|\hat{\rho}|\epsilon_2\rangle =
 -g_2\langle\epsilon_1|\hat{\rho}|\epsilon_1\rangle 
 +g_2^-\langle 0|\hat{\rho}|0\rangle 
 +\omega\Delta_2\sin\theta , \\
 && \partial_t\left(\Delta_1+i\Delta_2\right)= 
 \frac{i\omega\sin\theta}{2}
 \left(\langle\epsilon_1|\hat{\rho}|\epsilon_1\rangle 
 -\langle\epsilon_2|\hat{\rho}|\epsilon_2\rangle \right) \nonumber\\
 && \quad +\left[
 i(\epsilon_{21}-\omega\cos\theta)-\frac{1}{2}(g_1+g_2)
 \right]
 \left(\Delta_1+i\Delta_2\right).
\end{eqnarray}
Due  to the normalization of the density operator, 
the first three equations are not independent from each other.
We can write 
\begin{eqnarray}
 && 
 \left(\begin{array}{c}
 1 \\
 \partial_t\langle \epsilon_1|\hat{\rho}|\epsilon_1\rangle \\
 \partial_t\langle \epsilon_2|\hat{\rho}|\epsilon_2\rangle \\
 \partial_t\Delta_1 \\ \partial_t\Delta_2\end{array}\right) 
 = {\cal L} 
 \left(\begin{array}{c}
 \langle 0|\hat{\rho}|0\rangle \\
 \langle \epsilon_1|\hat{\rho}|\epsilon_1\rangle \\
 \langle \epsilon_2|\hat{\rho}|\epsilon_2\rangle \\
 \Delta_1 \\ \Delta_2\end{array}\right), \label{liouville}
\end{eqnarray}
where 
\begin{equation}
 {\cal L}=\left(\begin{array}{ccccc}
 1 & 1 & 1 & 0 & 0 \\
 g_1^- & -g_1 & 0 & 0 & -\omega\sin\theta \\
 g_2^- & 0 & -g_2 & 0 & \omega\sin\theta \\
 0 & 0 & 0 & -g & -\tilde{\epsilon} \\
 0 & \frac{\omega}{2}\sin\theta & -\frac{\omega}{2}\sin\theta & 
 \tilde{\epsilon} & -g 
 \end{array}\right). \label{calL}
\end{equation}
Here we use the notation 
$g=(g_1+g_2)/2$ and $\tilde{\epsilon}=\epsilon_{21}-\omega\cos\theta$.
Since each component of ${\cal L}$ is independent of $t$, 
the stationary solution is obtained by neglecting
the derivative on the left-hand side of Eq.~(\ref{liouville}).
Solving the equation algebraically, we obtain at the stationary limit 
\begin{eqnarray}
 && \langle 0|\hat{\rho}(t)|0\rangle \to \rho_0, \\
 && \langle\epsilon_1(t)|\hat{\rho}(t)|\epsilon_1(t)\rangle
 \to \frac{g_1^-}{g_1}\rho_0-\frac{\omega\sin\theta}{g_1}\Delta_0, \\
 && \langle\epsilon_2(t)|\hat{\rho}(t)|\epsilon_2(t)\rangle
 \to \frac{g_2^-}{g_2}\rho_0+\frac{\omega\sin\theta}{g_2}\Delta_0, \\
 && \Delta_2(t) \to \Delta_0,
\end{eqnarray}
where
\begin{equation}
 \Delta_0 = 
 -\frac{\omega\sin\theta}{2G}
 \left(\frac{g_2^{-}}{g_2}-\frac{g_1^{-}}{g_1}\right)\frac{1}{Z},
\end{equation}
and $\rho_0$ is given in Eq.~(\ref{rho0}). 
The heat flux in Eq.~(\ref{heat}) is expressed by 
using the above relations as 
\begin{eqnarray}
 \dot{Q}(t) 
 &=& 
 -\epsilon_{10}\left(g_1\langle\epsilon_1(t)|\hat{\rho}(t)|\epsilon_1(t)\rangle
 -g_1^-\langle 0|\hat{\rho}(t)|0\rangle\right) \nonumber\\
 &&-\epsilon_{20}\left(g_2\langle\epsilon_2(t)|\hat{\rho}(t)|\epsilon_2(t)
 \rangle
 -g_2^-\langle 0|\hat{\rho}(t)|0\rangle\right). \nonumber\\
\end{eqnarray}
By using Eqs.(\ref{g1})--(\ref{g2m}), 
we decompose this expression into two parts 
to write Eqs.~(\ref{qc}) and (\ref{qh}).
In a similar way, the work done by the system can be calculated as 
\begin{eqnarray}
 \dot{W}(t) = -{\rm Tr}\left[\hat{\rho}(t)\partial_t\hat{H}(t)\right] 
 = -2\lambda\omega \Delta_2(t),
\end{eqnarray}
which leads to the expression of the power in Eq.~(\ref{p}).
It is reasonable to find that the off-diagonal component 
$\langle\epsilon_1(t)|\hat{\rho}(t)|\epsilon_2(t)\rangle$
drives the system to make a finite amount of the work.

\section{Density operator at stationary}
\label{app-rho}

At the stationary limit, the density operator is written as 
\begin{eqnarray}
 \hat{\rho}(t) &=& \sum_{n=0}^2|\epsilon_n(t)\rangle
 \langle \epsilon_n(t)|\hat{\rho}(t)|\epsilon_n(t)\rangle
 \langle\epsilon_n(t)| \nonumber\\
 && 
 +|\epsilon_1(t)\rangle e^{i\omega t}\left(\Delta_1+i\Delta_2\right)
 \langle\epsilon_2(t)| \nonumber\\
 && 
 +|\epsilon_2(t)\rangle e^{-i\omega t}\left(\Delta_1-i\Delta_2\right)
 \langle\epsilon_1(t)|.
\end{eqnarray}
We diagonalize this operator as 
$\hat{\rho}(t)=\sum_n p_n|\rho_n(t)\rangle\langle\rho_n(t)|$.
The eigenvalues are given by 
\begin{eqnarray}
 && p_0=\rho_0,\\
 && p_1=\frac{1+\cos\Theta}{2\cos\Theta}\rho_1
 -\frac{1-\cos\Theta}{2\cos\Theta}\rho_2,\\
 && p_2=-\frac{1-\cos\Theta}{2\cos\Theta}\rho_1
 +\frac{1+\cos\Theta}{2\cos\Theta}\rho_2,
\end{eqnarray}
where $\rho_n=\langle \epsilon_n|\hat{\rho}|\epsilon_n\rangle$ and 
\begin{equation}
 \tan\Theta=\frac{|\Delta_1+i\Delta_2|}{\frac{1}{2}(\rho_2-\rho_1)}.
\end{equation}
At the stationary limit, these eigenvalues are independent of $t$.

The corresponding eigenstates are given by
\begin{eqnarray}
 && |\rho_0(t)\rangle=|0\rangle,\\
 && |\rho_1(t)\rangle=|\epsilon_1(t)\rangle\cos\frac{\Theta}{2}
 -|\epsilon_2(t)\rangle e^{-i\Phi(t)}\sin\frac{\Theta}{2},\\
 && |\rho_2(t)\rangle=|\epsilon_1(t)\rangle e^{i\Phi(t)}\sin\frac{\Theta}{2}
 +|\epsilon_2(t)\rangle \cos\frac{\Theta}{2},
\end{eqnarray}
where
\begin{equation}
 \Phi(t)=\omega t+{\rm arg} (\Delta_1+i\Delta_2).
\end{equation}

\section{Power fluctuation}
\label{app-tur}

The higher-order correlations of heat flows can be calculated by 
the introduction of the counting field~\cite{Kalaee21,Menczel21}.
The dissipator is modified as 
\begin{eqnarray}
 &&\hat{D}^\chi_\alpha[\hat{\rho}] = \sum_{\epsilon}\gamma_{\alpha}(\epsilon)
 \biggl[e^{-\chi_\alpha\epsilon}
 \hat{L}_{\alpha}^{\epsilon}(t)\hat{\rho}(\hat{L}_{\alpha}^{\epsilon}(t))^\dag
 \nonumber\\
 && -\frac{1}{2}\left( 
 (\hat{L}_{\alpha}^{\epsilon}(t))^\dag \hat{L}_{\alpha}^{\epsilon}(t)\hat{\rho}
 +\hat{\rho}(\hat{L}_{\alpha}^{\epsilon}(t))^\dag \hat{L}_{\alpha}^{\epsilon}(t)
 \right)
 \biggr]. \label{dissipatorchi}
\end{eqnarray}
$\chi_{\alpha}$ represents the counting field.
The average power is calculated by setting $\chi_{\rm c}=\chi_{\rm h}=\chi$ as 
\begin{equation}
 P=\lim_{T\to\infty}\frac{1}{T}
 \left.\frac{\partial}{\partial\chi}{\rm Tr}\,\hat{\rho}^\chi(T)
 \right|_{\chi=0}.
\end{equation}
In the same way, the variance of the power is given by 
\begin{equation}
 {\rm var}\,P=\lim_{T\to\infty}\frac{1}{T}
 \left[
 \left.\left(\frac{\partial}{\partial\chi}\right)^2
 {\rm Tr}\,\hat{\rho}^\chi(T)\right|_{\chi=0}
 -(PT)^2\right]. \label{varp}
\end{equation}

The GKLS equation with the dissipator in Eq.~(\ref{dissipatorchi}) 
is solved perturbatively.
The density operator and the dissipator are expanded with respect to 
the counting field $\chi_\alpha$ as 
$\hat{\rho}^\chi(t)=\hat{\rho}^\chi(t)+\hat{\rho}_1^{\chi}(t)
+\hat{\rho}_2^{\chi}(t)+\cdots$
and 
$\hat{D}_\alpha^{\chi}[\hat{\rho}]=\hat{D}_\alpha[\hat{\rho}]
+\hat{D}_{\alpha 1}^{\chi}[\hat{\rho}]
+\hat{D}_{\alpha 2}^{\chi}[\hat{\rho}(t)]+\cdots$, respectively.
Taking the trace of the GKLS equation at first order in $\chi$, we obtain
\begin{equation}
 \partial_t{\rm Tr}\,\hat{\rho}_1^{\chi}(t) = \sum_{\alpha}{\rm Tr}\, 
 \hat{D}_{\alpha 1}^{\chi}[\hat{\rho}(t)]
 = \sum_{\alpha}
 \chi_\alpha  {\rm Tr}\, \hat{J}_\alpha(t)\hat{\rho}(t), \label{firsttrrho}
\end{equation}
where 
\begin{equation}
 \hat{J}_\alpha(t)=-\sum_\epsilon \epsilon\gamma_\alpha(\epsilon)
  (\hat{L}_{\alpha}^{\epsilon}(t))^\dag\hat{L}_{\alpha}^{\epsilon}(t).
\end{equation}
$\hat{J}_\alpha(t)$ is interpreted as the current operator, which justifies 
the use of the dissipator in Eq.~(\ref{dissipatorchi}).
Then, by setting $\chi_{\alpha}=\chi$ and by using the stationary solution, 
we obtain
\begin{equation}
 \partial_t{\rm Tr}\,\hat{\rho}_1^\chi(t) \to \chi P.
\end{equation}
This result is consistent with that in Eq.~(\ref{W}).

Next, we consider the trace of the GKLS equation at second order given by 
\begin{equation}
 \partial_t{\rm Tr}\,\hat{\rho}_2^{\chi}(t) = 
 \sum_{\alpha}{\rm Tr}\, \hat{D}_{\alpha 1}^{\chi}[\hat{\rho}_1^\chi(t)]
 +\sum_{\alpha}{\rm Tr}\, \hat{D}_{\alpha 2}^{\chi}[\hat{\rho}(t)]. \label{2ndorder}
\end{equation}
In order to solve the equation, we need to find the explicit form 
of $\hat{\rho}_1^\chi$, not of ${\rm Tr}\,\hat{\rho}_1^\chi$.
The GKLS equation at first order in $\chi$ is given by
\begin{eqnarray}
 && {\cal L}\left(\begin{array}{c}
 \langle 0|\hat{\rho}_1^\chi|0\rangle \\
 \langle \epsilon_1|\hat{\rho}_1^\chi|\epsilon_1\rangle \\
 \langle \epsilon_2|\hat{\rho}_1^\chi|\epsilon_2\rangle \\
 \Delta_1^\chi \\
 \Delta_2^\chi 
 \end{array}\right) \nonumber\\
 &=& \left(\begin{array}{c}
 {\rm Tr}\,\hat{\rho}_1^\chi \\
 \partial_t\langle \epsilon_1|\hat{\rho}_1^\chi|\epsilon_1\rangle
 -\sum_\alpha\langle\epsilon_1|\hat{D}_{\alpha 1}^\chi[\hat{\rho}]|\epsilon_1
 \rangle \\
 \partial_t\langle \epsilon_2|\hat{\rho}_1^\chi|\epsilon_2\rangle
 -\sum_\alpha\langle\epsilon_1|\hat{D}_{\alpha 1}^\chi[\hat{\rho}]|\epsilon_1
 \rangle \\
 \partial_t\Delta_1^\chi \\
 \partial_t\Delta_2^\chi 
 \end{array}\right), \label{firstrho}
\end{eqnarray}
where ${\cal L}$ is given in Eq.~(\ref{calL}).
As we have shown above, $\hat{\rho}_1^\chi$ is a linear function of $t$.
This means that the first component of the vector on the right-hand side 
of Eq.~(\ref{firstrho}) is proportional to $t$ and 
the other components are independent of $t$.
By using Eqs.~(\ref{liouville}) and (\ref{firsttrrho}),
we can write the solution as 
\begin{eqnarray}
 &&\left(\begin{array}{c}
 \langle 0|\hat{\rho}_1^\chi|0\rangle \\
 \langle \epsilon_1|\hat{\rho}_1^\chi|\epsilon_1\rangle \\
 \langle \epsilon_2|\hat{\rho}_1^\chi|\epsilon_2\rangle \\
 \Delta_1^\chi \\
 \Delta_2^\chi 
 \end{array}\right) \nonumber\\
 &=&t\left(\begin{array}{c}
 \langle 0|\hat{\rho}|0\rangle \\
 \langle \epsilon_1|\hat{\rho}|\epsilon_1\rangle \\
 \langle \epsilon_2|\hat{\rho}|\epsilon_2\rangle \\
 \Delta_1 \\
 \Delta_2 
 \end{array}\right) \sum_\alpha{\rm Tr}\,\hat{D}_{\alpha 1}^\chi[\hat{\rho}]
 +{\cal L}^{-1}\sum_\alpha|\psi_\alpha^\chi\rangle,  \nonumber\\
\end{eqnarray}
where 
\begin{equation}
 |\psi_\alpha^\chi\rangle=\left(\begin{array}{c}
 0  \\
 \langle \epsilon_1|\hat{\rho}|\epsilon_1\rangle{\rm Tr}\,
 \hat{D}_{\alpha 1}^{\chi}[\hat{\rho}]-\langle \epsilon_1|
 \hat{D}_{\alpha 1}^\chi[\hat{\rho}]|\epsilon_1\rangle \\
 \langle \epsilon_1|\hat{\rho}|\epsilon_1\rangle{\rm Tr}\,
 \hat{D}_{\alpha 1}^{\chi}[\hat{\rho}]-\langle \epsilon_1|
 \hat{D}_{\alpha 1}^\chi[\hat{\rho}]|\epsilon_1\rangle \\
 \Delta_1{\rm Tr}\,
 \hat{D}_{\alpha 1}^{\chi}[\hat{\rho}] \\
 \Delta_2{\rm Tr}\,
 \hat{D}_{\alpha 1}^{\chi}[\hat{\rho}]
 \end{array}\right).
\end{equation}
This solution is inserted into Eq.~(\ref{2ndorder}).
${\rm Tr}\,\hat{\rho}_2^{\chi}(t)$ is a quadratic function of $t$.
The $t^2$ term represents the disconnected part of the fluctuation and
is canceled out by subtracting the square of the average 
as in Eq.~(\ref{varp}).
We obtain the expression of the variance given in the main text.

\end{document}